\documentclass[aps,prl,twocolumn]{revtex4}
\usepackage{amssymb}
\usepackage{amsmath}
\usepackage{bm}
\usepackage{graphicx}
\usepackage{epsfig,amsmath}

\begin{document}

\title{Reconnection of vortex filaments and Kolmogorov spectrum}
\author{Sergey K. Nemirovskii\thanks{%
email address: nemir@itp.nsc.ru}}
\affiliation{Institute of Thermophysics, Lavrentyev ave, 1, 630090, Novosibirsk, Russia
and Novosibirsk State University, Novosibirsk}
\date{\today }

\begin{abstract}
The energy spectrum of the 3D velocity field, induced by collapsing vortex
filaments is studied. One of the aims of this work is to clarify the
appearance of the Kolmogorov type energy spectrum $E(k)\varpropto k^{-5/3}$,
observed in many numerical works on discrete vortex tubes (quantized vortex
filaments in quantum fluids). Usually, explaining classical turbulent
properties of quantum turbulence, the model of vortex bundles, is used. This
model is necessary to mimic the vortex stretching, which is responsible for
the energy transfer in classical turbulence. In our consideration we do not
appeal to the possible "bundle arrangement" but explore alternative idea
that the turbulent spectra appear from singular solution, which describe the
collapsing line at moments of reconnection. One more aim is related to an
important and intensively discussed topic - a role of hydrodynamic collapse
in the formation of turbulent spectra. We demonstrated that the specific
vortex filament configuration generated the spectrum $E(k)$ close to the
Kolmogorov dependence and discussed the reason for this as well as the
reason for deviation. We also discuss the obtained results from point of
view of the both classical and quantum turbulence.
\end{abstract}

\maketitle
\affiliation{Institute of Thermophysics, Lavrentyev ave, 1, 630090, Novosibirsk, Russia
and Novosibirsk State University, Novosibirsk Russia}

\begin{figure}[tbph]
\includegraphics[width=6cm]{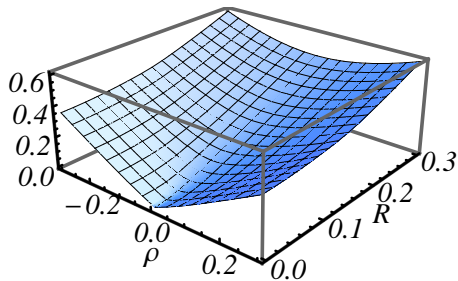}
\caption{(Color online) Quantity $D(\protect\rho ,R)$ - the phase function
in coordinates $\protect\rho ,R$. }
\label{3Ddistance}
\end{figure}

\emph{Background}.---We discuss the possibility of realization of the
Kolmogorov type energy spectrum $E(\mathbf{k})\varpropto \,k^{-5/3}$ of the
3D velocity field, produced by the vortex filament, collapsing towards
reconnection.The first motivation of this work is related to the problem of
modeling classical turbulence with a set of chaotic vortex filaments. This
idea has been discussed for quite a long time.(for details see, e.g. \cite%
{Chorin1994}-\cite{Nemirovskii2013}). In classical fluids thin vortex tubes
do not exist because they spread due to viscosity, so the concept of vortex
filaments should just be considered as a model. Quantum fluids, where the
vortex filaments are real objects, give an excellent opportunity for
developing the study of the question of whether the dynamics of a set of
vortex lines is able to reproduce (at least partially) the properties of
real hydrodynamic turbulence. \newline
Among various arguments supporting the idea of quasi-classic behaviour of
quantum turbulence, the strongest one is the $k$-dependence of the spectra
of energy $E(k)$ obtained in numerical simulations and experiments. There
are many works, which demonstrate dependence of $E(k)$ close to the
Kolmogorov law$\,E(k)\varpropto k^{-5/3}$. These are works, based on the
both vortex filament method \cite{Araki2002,Kivotides2002,Kivotides2001c},
and works using the Gross-Pitaevskii equation \cite{Nore1997}-\cite{Sasa2011}%
. The most common view of quasi-classical turbulence is the model of vortex
bundles. The point is that the quantized vortices have the fixed core
radius, so they do not possess the very important property of classical
turbulence -- stretching of vortex tubes with decrease of the core size. The
latter is responsible for the turbulence energy cascade from the large
scales to the small scales. Collections of near-parallel quantized vortices
(vortex bundles) do possess this property, so the idea that the
quasi-classical turbulence in quantum fluids is realized via vortex bundles
of different sizes and intensities (number of threads ) seems quite natural.
Meanwhile, a conception of the bundle structure is vague and up to now it
has not been definitely confirmed. It is unclear how the bundles can
spontaneously appear (at low temperature, when the coupling with normal
component is small). Moreover, even if they are prepared artificially, they
are extremely unstable (see, \cite{Nemirovskii2013},\cite{Volovik2004}),
they easily can be destroyed in result of reconnection either between of the
neighboring threads or in collisions with the other bundles, with the
forming of the \textquotedblleft bridging\textquotedblright . Therefore it
quite tempting to find another alternative mechanism of appearing of the
Kolmogorov type spectrum, and we offer the collapsing vortex filaments as a
candidate for this purpose.\newline
The second motivation is related to other important and intensively
discussed topic - a role of hydrodynamic collapse in formation of turbulent
spectra.(see e.g. \cite{Kuznetsov2000}, \cite{Kerr2013}). The striking
examples of such type spectra are the Phillips spectrum for water-wind
waves, created by white caps -- wedges of water surface of water surface or
the Kadomtsev-Petviashvili spectrum for acoustic turbulence created by
shocks \cite{Kuznetsov2000}. In the vortex filament theory the singularity
formation in a finite time arises due to approach of interacting vortex
filaments.\textit{\ }The result of this approach is appearing of very acute
kink, and energy of interaction between closely located parts can
essentially exceed contributions from a smooth elements of lines.\newline
In the work we introduce the general method for calculation of the energy
spectrum via the vortex line configuration, then we choose analytic relation
for the shape of kink, and conduct the mixed analyitic and numerical
evaluation of $E(\mathbf{k})$. We demonstrated that the spectrum $E(k)$ is
very close to the Kolmogorov dependence $\varpropto k^{-5/3}$, and discuss
the reason of this as well as the reason of deviation.\newline
\emph{Calculation of spectrum}.---The formal relation, allowing the calculation of $E(%
\mathbf{k})={\rho }_{s}\mathbf{v}_{\mathbf{k}}\mathbf{v}_{-\mathbf{k}}$ $/2={%
\rho }_{s}\bm{\omega }_{\mathbf{k}}\bm{\omega }_{-\mathbf{k}}/2k^{2}$ ($%
\mathbf{kv}_{\mathbf{k}}=0$ due to incompressibility) via the vortex line
configuration $\{\mathbf{s}(\xi )\}$, can be written as follows (see \cite%
{Nemirovskii1998},\cite{Nemirovskii2013a})%
\begin{equation}
E(\mathbf{k})=\frac{{\rho }_{s}{\kappa }^{2}}{16\pi ^{3}k^{2}}\oint \oint
\mathbf{s}^{\prime }(\xi _{1})\mathbf{s}^{\prime }(\xi _{2})d{\xi }_{1}d{\xi
}_{2}e^{i\mathbf{k(s}(\xi _{1})-\mathbf{s}(\xi _{2}))}.  \label{E}
\end{equation}%
Here $\mathbf{s}(\xi )=\bigcup \mathbf{s}_{i}(\xi _{i})$ is unification of
lines $\mathbf{s}_{i}(\xi _{i})$ where $\mathbf{s}_{i}(\xi _{i})$ describes
the $i-$vortex line position parameterized by the label variable $\xi _{i}$,
$\mathbf{s}_{i}^{\prime }(\xi _{i})$ denotes the derivative with respect to
variable $\xi _{i}$ (the tangent vector), and $\int\nolimits_{C}$ $%
=\int\nolimits_{C}\sum\nolimits_{j}$. In the isotropic case, the spectral
density depends on the absolute value of the wave number $k$. Integration
over the solid angle leads to the formula (see,\cite{Nemirovskii2013a}-\cite%
{Kondaurova2005}):
\begin{equation}
E(k)=\frac{\rho _{s}\kappa ^{2}}{(2\pi )^{2}}\oint \oint \mathbf{s}^{\prime
}(\xi _{1})\mathbf{s}^{\prime }(\xi _{2})d\xi _{1}d\xi _{2}\frac{\sin
(k\left\vert \mathbf{s}(\xi _{1})-\mathbf{s}(\xi _{2})\right\vert )}{%
k\left\vert \mathbf{s}(\xi _{1})-\mathbf{s}(\xi _{2})\right\vert }.
\label{E(k) spherical single}
\end{equation}%
For anisotropic situations, formula (\ref{E(k) spherical single}) is
understood as the angular average. Thus, for calculation of the energy
spectrum $E(k)$ of the 3D velocity field, induced by the collapsing vortex
filament we need to know an exact configuration $\{\mathbf{s}(\xi )\}$ of
vortex lines.\newline
\emph{Shape of kink}.---Despite the huge number of works devoted to dynamics
of collapsing lines both in classic and quantum fluids \cite{Siggia1985}-%
\cite{Boue2013} (this list is far not full) the exact solution $\mathbf{s}%
(\xi )$ for the shape of curves has not obtained up to now The main results
were obtained by different approaches, combining analytical and numerical
methods, such as the local induction approximation and full Biot-Savart law,
and also Nonlinear Schr\"{o}dinger equation for vortices in Bose- Einstein
condensate.
\begin{figure}[h]
\includegraphics[width=6.5cm]{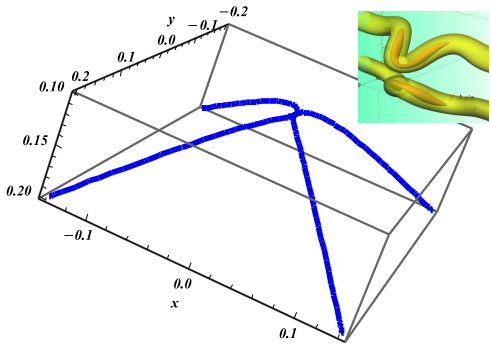}
\caption{(Color online) The touchng quasi-hyperbolae describing the
collapsing lines (see Eq. (\protect\ref{curves}) ) obtained in \protect\cite%
{Boue2013}. In the inset we set (as an example) the kinks on the
anti-parallel collapsing vortex tubes obtained in numerical simulation
\protect\cite{Bustamante2008} }
\label{hyperbolas}
\end{figure}
Qualitatively the results of these investigations are quite similar and can
be described as follows. Due to long range interaction in the Biot-Savart
integral, the initially arbitrarily oriented vortices, when they approach
each other, they start by reorienting their close segments so as to bring
them into an antiparallel position. Further, cusps may appear on the
approaching segments of two vortex lines. The curvature of these cusps may
be so large that the self-induced velocity of each perturbation overcomes
the repulsion from the adjoining vortex line. Further the cusps grow,
approach each other closer increasing their curvature and correspondingly
their self induced velocities and this process is repeated faster and
faster. It is important that this process grows explosively, since the
distance between the two perturbed segments, $\Delta $, decreases according
to the relation $\;\Delta \sim (t^{\ast }-t)^{1/2}$, where $t^{\ast }$ is
some quantity depending on the relevant parameters and initial conditions.
Thus, in a finite time the vortex lines collapse. Asymptotic lines are two
hyperbolic curves lying on opposite sides of the pyramid (see e.g. \cite%
{Waele1994},\cite{Tebbs2011}). However, in recent study, \cite{Boue2013} it
was shown that the curves are not exact hyperbolas, but slightly different
lines (the authors called these curves as quasi-hyperbolae) of type $h(\xi )=
$ $\sqrt{a^{2}\xi ^{2}/(a^{2}+\xi ^{2})+a^{2}+\xi ^{2}}$, and that they lie
not in the planes of the pyramid sides, but on the curved surfaces, bent
inwards. In the moments just before the collapse, when the vortex cores are
nearly touch each other, the very acute kink appears. This curves may be
written in parametric form (cf. formula (16) of \cite{Boue2013})
\begin{equation}
\mathbf{s}_{1,2}(\xi )=\left[ \pm \left( h(\xi )-c\right) ,\ \pm \xi ,\
\left( h(h(\xi )-b)\right) \right]   \label{curves}
\end{equation}%
The described configuration is shown in Fig. \ref{hyperbolas}. The signs are
chosen so that $\mathbf{s}_{1}^{\prime }(0)\cdot \mathbf{s}_{2}^{\prime
}(0)=-1$ (the vortices are antiparallel). Quantity $a$ is of the order of
the curvature radius on the tip of the kink of curve, quantity $b$ (related
to $\ a$, see \cite{Boue2013}) is responsible for bending of the surfaces,
on which the quasi-hyperbolae\ lie. Quantity $c$ is also of the order of $a$
is responsible for closeness of the filaments. All this three quantities are
smaller of intervortex space $\delta =\mathcal{L}^{-1/2}$ (where $\mathcal{L}
$ is the vortex line density). This vision is consistent with the results of
numerous numerical works, studying the collapse of vortex lines (see, e.g.,
\cite{Kerr2013},\cite{Bustamante2008} and references therein, the decisive
picture obtained in \cite{Bustamante2008} is shown in the inset of Fig. \ref%
{hyperbolas}).\newline
\emph{Numerical results}.---In the left graphic of Fig. \ref%
{spectrumNumerical} we presented the results of numerical calculation of
spectrum $E(k)$ on the base formula (\ref{E(k) spherical single}) (without
prefactor before integral) using a configuration $\{\mathbf{s}(\xi )\}$ of
vortex lines, described by (\ref{curves}). We chose the following
parameters: $a=0.1,b=0.09,c=0.1$ (the case $a\approx c$ corresponds to
nearly touching curves). It is seen that in interval of wave numbers $k$
between $1\div 50$ the slope of $E(k)$ is indeed close to $-5/3$. We discuss
the origin of this in the following paragrph.\newline
\begin{figure}[tbph]
\includegraphics[width=8.5cm]{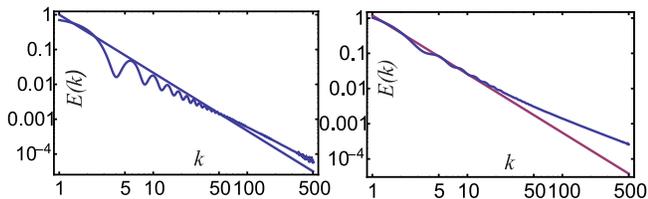}
\caption{(Color online) Right. The spectrum $E(k)$, obtained numerically on
the base formula (\protect\ref{E(k) spherical single}. The straight line has
a slope -5/3. Left. The same spectrum obtained on the basis of procedure
described in the paragraph "\emph{Analytic consideration}" (Eq. (\protect\ref%
{oneDintegral})). }
\label{spectrumNumerical}
\end{figure}
\emph{Analytic consideration}.---Because of rapidly oscillating function,
the evaluation of integral (\ref{E(k) spherical single}) is difficult, even
numerically. In addition, numerical results obscure underlying physics,
therefore we intend to perform analytical study, at least as far as
possible. The integral (\ref{E(k) spherical single}) can be approximately
evaluated for large $k$ using the method of asymptotic expansion \cite%
{Fedoryuk1977}. When $k$ is large the function $\sin (k\left\vert \mathbf{s}%
(\xi _{1})-\mathbf{s}(\xi _{2})\right\vert )$ is rapidly varying function,
therefore the main contribution into integral comes from points of minimal
value of the separation funcion between points of the curves $D(\xi _{1},\xi
_{2})=$ $\left\vert \mathbf{s}(\xi _{1})-\mathbf{s}(\xi _{2})\right\vert $.
This is enhanced by the fact that the distance is included in the
denominator in the integrand of (\ref{E(k) spherical single}). Thus, the
behaviour of the phase function $D(\xi _{1},\xi _{2})$ near minimum is
crucial for value of integral and for its $k$-dependence. Let us study the
phase function $D(\xi _{1},\xi _{2})$ for the vortex configuration described
by Eq. (\ref{curves}) just before collapse when $c\approx a$. It is
convenient to introduce variables $\rho =\xi _{1}-\xi _{2}$ and $R=(\xi
_{1}+\xi _{2})/2$ and recast the double integral $\int\nolimits_{C}\int%
\nolimits_{C}d\xi _{1}d\xi _{2}$ as multiple integral $\int dR\int d\rho $ \
in the domain bounded by lines $\rho =2R$ and $\rho =-2R$. The upper limit
for $R$ is not essential, since the integral gains the main contribution
from vicinity of point $R=0$. Let's consider the behaviour of function $%
D(\rho ,R)$. It is depicted in Fig \ref{3Ddistance} (in the extended domain)
At the beginning of coordinates $\rho =0,R=0$ function $D(\rho ,R)=0$. The
important feature of function $D(\rho ,R)$ is its behaviour of it near
points $\rho =0$, (for different $R$), that is median part of domain,
arising from equidistant ($\xi _{1}=\xi _{2}$) points of the touching vortex
filaments. For fixed $R$ (perpendicular to the median direction) the
functions $D(\rho ,R=const)$ are approximated by pieces of parabolas $%
\varpropto $ $\rho ^{2}$, then transferring into linear funcion $\varpropto
\left\vert \rho \right\vert $ for $\rho \gtrsim a$, with the same slope for
all $R$. Thus, all points of median are points of local minimum, and $\left.
\partial D/\partial \rho \right\vert _{\rho =0}=0$ for all $R$. Another
imporant feature of the phase function $D(\rho ,R)$ is its dependence on $R$
along the median $\rho =0$.
\begin{figure}[tbph]
\includegraphics[width=8.5cm]{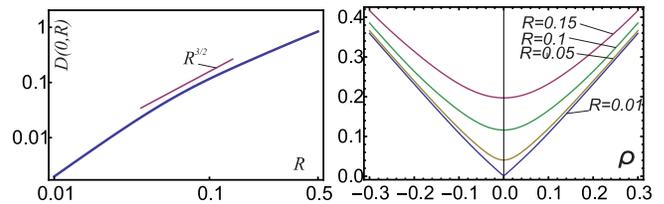}
\caption{(Color online) Left. Function $D(0,R)$, distance along the median
in the log-log coordinates. \ Segment of straight line has slope 3/2. Right.
The slices of the phase function $D(\protect\rho ,R)$ for different $R$.}
\label{medianlog}
\end{figure}
Fuction $D(0,R)$ is depicted in logarithmic coordinated in Fig. \ref%
{medianlog}. It is seen behaves as $\varpropto R^{2}$ then transferring into
$\varpropto R^{1}$ for $R\sim a$ (crossover region). Thus, we have\
complicated case, when point $(R=0,\rho =0)$ is simultaneously both a corner
of domain restricted by curves $\rho =2R$ and $\rho =-2R$, and a stationary
point (minimum), i.e. $\nabla D(\rho ,R)=0$.\newline
To move further we pass to polar coordinates $R,\theta $, then and
integrating over angle $\theta $ we obtain asymptotic expansion over $1/k$.
The leading term has the form%
\begin{equation}
E(k)=\int\limits_{0}dR\sqrt{\frac{\pi }{k\left. \partial ^{2}D/\partial \rho
^{2}\right\vert _{\rho =0}}}\left. \frac{\sin (kD(\rho ,R))}{kD(\rho ,R)}%
\right\vert _{\rho =0}  \label{oneDintegral}
\end{equation}%
We used hear that integration over $\theta $ is alike integration over $%
d\rho $, namely $d\rho =Rd\theta $ and the median curve $\rho =0$ is the
line where function $D(\rho ,R=const)$ has a local minimum $\left. \partial
D/\partial \rho \right\vert _{\rho =0}=0$. Therefore the integartion over $%
\theta $ can be carry out \ by the use of method of stationary phase, which
gives \ref{oneDintegral}. Calulating the integral in vicinity of staionary
point we neglected the slowly changing function $\mathbf{s}^{\prime }(\xi
_{1})\cdot \mathbf{s}^{\prime }(\xi _{2})$, putting it to be equal to $-1$
(we recall, that the lines are antiparallel). Additionally, we take $\sin
(kD(\rho ,R))$ as a imaginary part of $\exp (ikD(\rho ,R)).$Thus, we reduced
the whole problem to evaluation of the 1D integral. In the right graphic of
Fig \ref{spectrumNumerical} we presented $E(k)$, calculated on the basis of
formula (\ref{oneDintegral}). First of all please note that spectrum
calculated with use of (\ref{oneDintegral}) very close to the spectrum
calculated on basis (\ref{E(k) spherical single}), this justifies the
approximated procedure, described above. Second, and more imortant fact is
that again in interval of wave numbers $k$ between $1\div 50$ the slope of $%
E(k)$ is close to $-5/3$.\newline
To understand an appearance of the $\approx k^{-5/3}$ dependence we appeal
to the so called Erdelyi lemma \cite{Erdelyi1955},which says that the
intergral $\int\nolimits_{0}x^{\beta -1}f(x)e^{i\lambda x^{\alpha }}dx$ \
with a smooth enough function $f(x)$ has an expansion in asymptotic series
as $\sum_{m}a_{m}\lambda ^{-\frac{m+\beta }{\alpha }}$ with the leading term
$\lambda ^{-\frac{\beta }{\alpha }}$. That, in particular, means that if we
took the collapsing filaments not to be quasi-hyperbolas but was pure
power-like functions $\mathbf{s}_{1,2}(\xi )=(\xi ,\pm \xi ^{3/2},0)$ ($3/2$
parabolas) and implemented the procedure discribed above, we would obtain
the spectrum had exact $E(k)\varpropto $ $k^{-5/3}$ form Coming back to
solution (\ref{curves}) and Fig. (\ref{medianlog}) we see that intervortex
distance (along the median $\xi _{1}=\xi _{2}$) is not $3/2$ parabola but it
is \ more sophysticated funtion which behaves as $\varpropto R^{2}$ then
transferring into $\varpropto R^{1}$ in the crossover region $\Delta $,
covering $1\div 1.5$ decades near quantity $a$. Therefore in the crossover
region where the quantity $D(0,R)$ is close to $R^{3/2}$, it should be
expected, that $E(k)$ is close to the Kolmogorov dependence $\varpropto
k^{-5/3}$ for the wave numbers $k$ of the order $2\pi /\Delta $,
which,.indeed, takes a place. The crossover region lies from the scale of
bend $a$, and scale where branches of hyperbolas tends to straght lines.
Actually it is close to size of the bridging kink on the curves and is of
the order of intervortex space $\delta =\mathcal{L}^{-1/2}$ (see the right
picture on Fig. \ref{hyperbolas}). In fact, the numerical works \cite%
{Araki2002}-\cite{Sasa2011} cited in the Introduction the authors obtained
the spectrum $E(k)\approx k^{-5/3}$ only for the wave numbers $k$, around $%
k\approx $ $2\pi /\delta $. \newline
\ \ \ \emph{Conclusion}.--- Coming back to the aims of work stated in the
Introduction we can suggest that the spectrum $E(k)$ close to the Kolmogorov
dependence $\varpropto k^{-5/3}$ , which was observed in many numerical
simulations on the dynamics of quantized vortex filaments \cite{Araki2002}-%
\cite{Sasa2011}, can appear from the reconnecting lines. Unfortunately,
because of lack of exact analytic solution for the configuration $\{\mathbf{s%
}(\xi )\}$ of the collapsing vortex filaments, the quantity $E(k)$ is
approximate and relies on the asymptotic solution (\ref{curves}). On the
other hand (as seen from the proposed analytical consideration) spectrum
depends on few features of collapsing line, such as as order of touching and
the crossover to smooth straight line. These features are universal and
observed in in many numerical simulations.\newline
Another, more delicate question, touched in the Introduction, concerns the
role of dynamics of discrete vortices in the physics of turbulence. On the
one hand, our results support the point of view on the role collapse in the
formation of turbulent spectra conducted in \cite{Kuznetsov2000}. On the
other hand, many unclear questions remain. In the Kolmogorov scenario the
spectrum $E(k)\varpropto $ $k^{-5/3}$ was the consequence of a $k$%
-independent energy cascade $P_{k}$ in the $k$ space. In the scheme based on
collapsing lines the energy cascade does not appear at all (at least in an
explicit form). Although due to dimension speculations something like $%
E(k)\varpropto (P_{k})^{2/3}$ should appear, but the question how it comes,
is unclear. It can be put forward an assumption that the collapse of lines,
which delivers energy into a tiny region near the point of collapse (then
this energy is burned in the process of full reconnection) plays the role of
the vortex stretching in the transfer of energy to small scales. But it is
just qualitative guess that is not supported by any quantitative
calculations. Another question concerns the interval of wave numbers where
the spectrum $E(k)$ $\approx k^{-5/3}$ is observed. It is regulated by the
curvature of the kink and intervortex space, so in reality this spectrum
covers maximum $1\div 1.5$ decades, while in real turbulence the Kolmogorov
dependence $\varpropto k^{-5/3}$ is observed for $3\div 4$ decades. Of
course, many other issues relating higher structure functions, the number of
reconnection, which is necessary to maintain a uniform spectrum, the exact
shape of the collapsing curves, etc. remain open, but these issues are
outside the framework of the presented work.\newline
The work was supported by 13-08-00673 from RFBR (Russian Foundation of
Fundamental Research).


\end{document}